\documentclass[twocolumn,amsmath,secnumarabic,amssymb, nobibnotes, aps, prd]{revtex4}

\usepackage{color}
\usepackage{pstricks}
\usepackage{graphicx}

\begin{document}

\title{Holographic Schwinger Effect}%

\author{Gordon W.~Semenoff}
\affiliation{ Department of Physics and
Astronomy, University of British Columbia, 6224 Agricultural Road,
Vancouver, British Columbia V6T 1Z1 }
\author{Konstantin Zarembo}
\affiliation{Nordita, Roslagstullsbacken 23, SE-106 91 Stockholm, Sweden and ITEP, Moscow, Russia.\\
 {\rm Preprint\# NSF-KITP-11-181; NORDITA-2011-73; 
Uppsala: UUITP-26/11}}
\begin{abstract}
We study tunneling pair creation of W-Bosons by an external electric field on the Coulomb branch of N=4 supersymmetric Yang-Mills theory.  We use AdS/CFT holography to find a generalization of Schwinger's formula for the pair production
rate
to the strong coupling, planar limit which includes the exchange of virtual massless particles to all orders.
We find that the pair creation formula has an upper critical electric field beyond which the process is no longer
exponentially suppressed. The value of the critical field is identical
to that which occurs in the Born-Infeld action of probe D3-branes in the $AdS_5\times S^5$ background.
\end{abstract}

\maketitle
One of the interesting attributes of string theory
is the existence of an upper critical electric field\cite{Fradkin:1985ys}\cite{Bachas:1992bh}.  Opposite electric charges
reside at the endpoints of open strings. An electric field pulls them in opposite directions.  When the field
exceeds the string tension, the barrier to stretching the string disappears and strings are unstable.
One might wonder, especially in light of AdS/CFT
holography, whether this phenomenon can also be visible in a quantum field theory.

In this Letter, we shall address
this issue in the quantum field theory for which holography is most firmly established, the super-conformal ${\cal N}=4$
Yang-Mills theory.
There, we can find an electric field by studying the theory on the Coulomb branch,
with gauge group U(N+1)
spontaneously broken by a vacuum
expectation value of the scalar fields to $U(1)\times U(N)$. We then
consider an electric field of the U(1) gauge theory.
It would act on the massive W-Bosons which
have U(1) charges $\pm g_{\rm YM} $ (the Yang-Mills coupling constant) and transform in the fundamental
representation of the residual gauge group U(N).
The W-Bosons are $\frac{1}{2}$-BPS particles
and form a short multiplet of the ${\cal N}=4$ supersymmetry algebra which contains scalar,
spinor and vector fields.
We will study a dynamical process, the  Schwinger effect of pair production for $W^\pm$
Bosons in a constant electric field.
\setcounter{footnote}{0}

Before we discuss holography, let us consider a simple
field theory argument as to why there could be an upper critical
electric field.
In order to become real particles, a virtual particle-antiparticle pair that are
created by a vacuum fluctuation must gain an energy equal to their combined rest masses $2m $ (we set $c=1=\hbar$).  This energy could be supplied by an electric field where they are pulled in opposite directions. Upon separating by a distance $d$, they gain energy $Ed$ and become physical particles when
$d\sim 2m /E$.  This process is tunneling through a barrier of height $\sim 2m $ and width  $\sim 2m /E$.  The
amplitude should therefore
be suppressed by an exponential of the product, $\sim \frac{m^2}{E}$.  Exponential suppression with this quantity in the exponent indeed appears in the formula
 for the tunneling probability which was
computed long ago by Schwinger \cite{Schwinger:1951nm} and is quoted in eq.~(\ref{schwinger}) below. We shall be interested in how Coulomb interactions would modify this effect.  With a Coulomb interaction added, the tunneling barrier has profile
$
 V_{\rm eff}(d)=2m-Ed-\frac{\alpha}{d}\,,
$
where $\alpha$ contains the electric charge.
\begin{figure}[t]
\centerline{\includegraphics[scale=0.7]{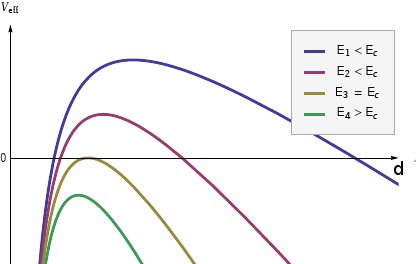}}
\caption{\label{astr}\small The effective potential of a created pair. }
\end{figure}
This barrier is shown in fig.~\ref{astr} for different values of the electric field. If the field is sufficiently small  it is positive in a certain range of distances, and the asymptotic region where the particles can be on-shell is separated from the origin, where the pair is created, by a potential barrier. Consequently at small fields the pair creation is a tunneling process and its amplitude is exponentially suppressed. However, when the field reaches the critical value, the potential is negative everywhere and pair creation does not require tunneling.  Its probability is no longer exponentially suppressed. The ${\cal N}=4$ Yang-Mills theory that we are interested in is indeed in a Coulomb phase and the
effective coupling of the planar limit of the theory at large 't Hooft coupling, $\lambda=g_{\rm YM}^2N$, deduced from
and AdS/CFT computation of the W-Boson Wilson loop for parallel lines in Ref.~\cite{Maldacena:1998im} is $\alpha=\frac{4\pi^2\sqrt{\lambda}}{\Gamma^4
 \left(\frac{1}{4}\right)}$.
An estimate of the critical field is
\begin{align}
 E_c\sim \frac{m^2}{\alpha}=\frac{\Gamma^4
 \left(\frac{1}{4}\right)m^2}{4\pi^2\sqrt{\lambda}}\approx 0.70\,\frac{2\pi m^2}{\sqrt{\lambda}}\,.
\label{emax00}\end{align}
which is remarkably similar to one which which we shall find in (\ref{emax0}) below  using holography.
This simple argument underestimates it by about 30\%.

In the IIB string theory, which is the holographic dual of ${\cal N}=4$ Yang Mills theory, the
Higgsing of U(N+1) to $U(1)\times U(N)$ is gotten by separating one D3-brane from a parallel
stack of N coincident D3-branes.
W-Bosons are open strings stretched between the separated D3-brane and the stack.
In the large N limit, the stack of D3-branes is
replaced  by the  $AdS_5\times S^5$ background with metric $dS^2=L^2(
r^2dx_\mu dx^\mu + \frac{dr^2}{r^2}+ d\Omega^2_5)$ and $N$ units of Ramond-Ramond 4-form flux.
The radius of curvature $L$ of the background is related to the Yang-Mills theory 't Hooft coupling
$\lambda= g_{\rm YM}^2N$ and the fundamental
 string scale $\ell_s=\sqrt{2\pi\alpha'}$ by $L^2= \sqrt{\lambda}\ell_s^2/2\pi $.
 The probe D3-brane  fills the four space-time
dimensions $x^\mu$ and sits at a fixed $AdS$ radius $r_0$ and at a point on $S^5$ (given by a unit 6-vector $\hat n$
in (\ref{wilson}) below).  In the large $N$ planar limit, where $N$ is set to infinity holding $\lambda$ fixed, and then
in the large $\lambda$ limit, the probe D3-brane can be treated as a classical object embedded in $AdS_5\times S^5$  \footnote{The probe brane limit which we employ here
 has been widely used for studying ${\cal N}=4$ Yang-Mills theory
  with additional fundamental representation fields \cite{Karch:2002sh}\cite{Karch:2007pd}.}.
 It has world-volume metric  $ds^2=
{L^2} {r_0^2}dx_\mu dx^\mu$
and a U(1) worldvolume gauge symmetry.  If
we turn on an electric field $E$ on the worldvolume of the probe brane,
it is described by the Dirac-Born-Infeld and Wess-Zumino actions
\footnote{We
normalize $E$ so that it is the Lorentz force on the W-Boson in its classical equation of motion,
$m\ddot x_\mu=F_{\mu\nu}\dot x_\nu$.
We hold this
Lorentz force fixed in the large $N$ limit.   },
 \begin{align}
S &=T_3\int d^4x\left[- \sqrt{-\det(g_{\mu\nu}-  \ell_s^2 F_{\mu\nu})}+\omega^{(4)}\right]
\label{dbi1} \\
&=\frac{  L^4 r_0^4}{2\pi\ell_s^4 g_s}\int d^4x \left[-
\sqrt{1-\frac{\ell_s^4}{L^4r_0^4}   E^2}+1\right]
\label{dbi2}\end{align}
Here, $T_3$ is the D3-brane tension, $\omega^{(4)} $ is the Ramond-Ramond 4-form and $g_s= g^2_{\rm YM}/4\pi$
is the closed string coupling constant.
For a flat brane on flat space, $g_{\mu\nu}$ in (\ref{dbi1}) would be replaced by the flat Minkowski space
metric diag$(-1,1,1,1)$  and (\ref{dbi1})
would indeed be singular for an electric field
where the argument of the square root vanishes,
$E_{\rm crit.}^{\rm flat}=\tfrac{1}{\ell_s^2}=\tfrac{1}{ 2\pi\alpha'}$, equal to the
flat space string tension, as expected.
In the AdS background, the metric in (\ref{dbi1}) has a warp factor,
the appropriate action is (\ref{dbi2}) and it also has a critical
field, $E_{\rm c}= L^2 r_0^2/\ell_s^2 $.  It is interesting to write this critical
field using Yang-Mills theory parameters.   The parameter $r_0$ is related to
the mass of the W-Boson, which is given by the energy of a fundamental string that is suspended between the probe
brane at $r=r_0$ and the Poincar\`{e} horizon at
$r=0$.  The worldsheet metric of such a string would be $d\sigma^2=L^2(- {r^2} dt^2+\frac{dr^2}{r^2})$
and the mass is
\begin{equation}\label{wbosonmass}
m = \frac{1}{\ell_s^2}\int_0^{r_0} dr \sqrt{-|g(t,r)|}=\frac{L^2}{\ell_s^2}\int_0^{r_0}dr=\frac{\sqrt{\lambda}r_0}{2\pi}
\end{equation}
Using (\ref{wbosonmass}) to eliminate $r_0$ from our expression for the critical electric field, we
conclude that
\begin{equation}
E_{\rm c}=\frac{2\pi m^2}{\sqrt{\lambda}}\label{emax0}\end{equation}
Duality  then indicates that the large N and large $\lambda$ limit of ${\cal N}=4$ Yang Mills theory
on its Coulomb branch
also has this critical field, which should be compared with (\ref{emax00}).
In the following, we will  examine the field theory manifestation of
this critical field in more detail.

Schwinger's formula \cite{Schwinger:1951nm} for the probability of production of charged
particle-antiparticle pairs, $P=1-e^{-\gamma v}$, where the contribution to $\gamma$ from a particle
with spin $j$ and mass $m$ in a constant electric field $E$ over space-time volume $v$ is,
\begin{equation}\label{schwinger}
 \gamma_j=\frac{(2j+1) E^2}{8\pi ^3}\sum_{n=1}^{\infty }\frac{\left(- 1\right)^{(n+1)(2j+1)}}{n^2} e^{-\frac{\pi
 m^2n}{| E|}}
\end{equation}
This formula applies to a weak coupling limit where radiative corrections from emission and reabsorption of virtual
particles are neglected.
It can be computed, as Schwinger originally did, from a proper time representation of the appropriate Feynman diagrams. Alternatively, it can be found (for a spin 0 particle) from the imaginary part of the Euclidean world-line path integral
\begin{eqnarray}
\label{svac0}
 \gamma_0=-\frac{2}{v}\Im
 \int_0^\infty\frac{dT}{T}\int \mathcal{D}x_\mu  ~
 e^{-\int_{0}^{1}d\tau \left[\frac{\dot{x}^2}{4T}
 +m^2T
-i  A_\mu  \dot x^\mu \right]}
 \end{eqnarray}
with periodic boundary conditions $x_\mu(\tau+1)=x_\mu(\tau)$. For a constant field,
$A_\mu = -\frac{1}{2}F_{\mu\nu}x_\nu$.
The non-zero components of $F_{\mu\nu}$ are
$F_{12}=-F_{21}=-iE$, the ``i'' a result of analytic continuation to Euclidean space and
we shall assume $E>0$.
As the exponentials in the summand in (\ref{schwinger}) suggest,
(\ref{svac0}) can be computed as a sum over instanton amplitudes for tunneling through the potential barrier
of pair creation~\cite{Affleck:1981bma}. In Euclidean space, the electric field acts as a magnetic field and
the instanton of the Euclidean world-line path integral is a cyclotron orbit of the charged particle, that is,
a circle trajectory.  The integer $n$ which is summed in (\ref{schwinger}) is the number of instantons, that is
the number of times
the particle traverses the circle. With the solution of the equation of motion
for $T$, $T=\tfrac{1}{2m}\sqrt{\int\dot x^2}$, and the circle Ans\"atz, $x_{0\mu}=R\hat r$
where $\hat r=\left( \cos(2\pi n\tau),\sin(2\pi n\tau),0,0\right)$, $n=1,2,\ldots$,
 the classical action is
\begin{align}S_{\rm cl(0)}=2\pi n Rm-\frac{2\pi n ER^2}{2}
\label{classical}\end{align}
The classical equation of motion for $x_\mu$ is solved when $R$ is adjusted
to an extremum of (\ref{classical}), $R=m/E$, which
is a maximum.  Fluctuations of $R$ are tachyonic and integrating them would
produce the factor of ``i'' which gives the Euclidean path
integral an imaginary part.
Substituting $R=m/E$ into (\ref{classical}) yields $S_{\rm cl(0)}=\tfrac{\pi m^2}{E}n$
which is identical to the exponent in Schwinger's formula, eq.~(\ref{schwinger}).
To determine the prefactor of the exponential in (\ref{schwinger}) it is
necessary to analyze fluctuations  about the
classical solution.  Ref.~\cite{Affleck:1981bma} showed how to get the prefactor
of the $n=1$ term by doing the quadratic integral over fluctuations.  Though it would be
desirable to do so here, for example, to understand the nature of the amplitude when exponential
suppression is absent, we will not
address this interesting problem in the present paper, but will reserve it for a more
detailed exposition elsewhere.

In the planar limit of ${\cal N}=4$ Yang-Mills theory, Schwinger's formula (\ref{schwinger}) applied to W-Bosons
would be modified in a number of ways.
It will have an overall factor of $N$ to reflect the number of W-Bosons and it
comes from the vacuum energy with one W-Boson loop.  Contributions with additional W-Boson loops are suppressed by
factors of $1/N$ and can be ignored in the large $N$ limit. As well, contributions from virtual  U(1) photons
 are proportional to $g_{\rm YM}^2=\lambda/N$ and are suppressed at large $N$.
 Interactions with the massless particles of the unbroken U(N) gauge theory are ignored in the weak coupling limit which produces (\ref{schwinger}), but must be included as, in the large $N$ limit, planar Feynman diagrams will contribute at all orders in  $\lambda$. For a scalar field in the W-Boson supermultiplet,
 these contributions (as well as the overall factor of $N$) can be taken into account
 by adding a Wilson loop amplitude
to the path integral (\ref{svac0}). The action in (\ref{svac0}) becomes
\begin{eqnarray}
 S  &=&
 \int_{0}^{1}d\tau \left[\frac{\dot{x}^2}{4T}
 +m^2T
+\frac{iF_{\mu\nu}}{2}x_\mu\dot{x}_\nu\right]
 -\ln W[x_\mu]
 \label{action01}
 \end{eqnarray}
 where $W[x_\mu]$ is the Wilson loop.
The appropriate quantity (in the large W-mass limit) is
\cite{Rey:1998ik} \cite{Maldacena:1998im} \cite{Drukker:1999zq}
\begin{equation}\label{wilson}
 W[x]=\left\langle \,{\rm Tr}\,{\cal P}e^{
 \int_{0}^{T}d\tau \left(i\dot{x}_\mu A_\mu +|\dot{x}|\hat n_I\Phi _I \right)
 } \right\rangle
\end{equation}
Here, $\hat n_I$ is a unit vector in the direction of the scalar field condensate $\left<\Phi_I\right>$.
The gauge field $A$ and scalar $\Phi^I$ transform in the adjoint representation of SU(N) and
the trace over
SU(N) indices is of order $N$.
The  path integral with action (\ref{action01}) is semi-classical when
the mass of the W-boson is large, $m^2>>E$.  We shall also consider strong coupling, $\lambda>>1$.
These limits are compatible with electric fields in the range $E\sim\tfrac{m^2}{\sqrt{\lambda}}$ where we expect to
find a critical field.
The conformal symmetry of ${\cal N}=4$ Yang-Mills theory implies that,
when evaluated on a circle, $x = R\hat r$, $W$ is a function of $mR$ and rotation symmetry implies
$\left.\tfrac{\delta}{\delta x^\mu}  W[x]\right|_{x={\rm R\hat r}}=\hat r_\mu \tfrac{d}{dR}  W $. Consequently, once the radius is adjusted
to an extremum of the action, now including the Wilson loop,
the circle is still a solution of the classical equation of motion derived from (\ref{action01}).
Moreover, for the infinite $mR$ limit, exact results for $W[{\rm circle}]$ \cite{Erickson:2000af}
\cite{Drukker:2000rr} \cite{Pestun:2007rz} and an expression for quadratic fluctuations
about a straight line  \cite{Polyakov:2000ti} \cite{Semenoff:2004qr} which can easily be adapted to a circle
are available.  Indeed, the known strong
coupling behavior for a circle (wrapped n times), $\ln W \sim n\sqrt{\lambda}$, combined with (\ref{classical}) would lead to a corrected
classical action $S_{\rm  cl(1)}=(\tfrac{\pi m^2}{ E}-\sqrt{\lambda})n $ which suggests a critical behavior at large $\lambda$ where $S_{\rm cl(1)}$ goes to zero and the
sum over $n$ in the Schwinger amplitude would no longer be exponentially suppressed.  However,
the computation of the Wilson loop we are using
is already specialized to infinite $mR\sim \tfrac{m^2}{E}$.  (A symptom of the problem is the fact that the critical $E$ that we
would estimate from the discussion above differs from (\ref{emax0}) by a factor of 2.)

To correctly estimate $\ln W$, we shall need the expectation value of the appropriate loop
with large but finite W-mass.  For this we return to the probe D3-brane placed at radius $r_0$ in $AdS_5$
and replace the action in  (\ref{classical}) by the disc amplitude
 for a string which intersects a probe D3-brane on the circle,
 $x(\tau)=R\hat r $ and couples to the electric field at the boundary of its worldsheet. In the large
 $\lambda$ limit, the string sigma model is semiclassical and the problem reduces to finding a disc of extremal
 area. The sigma model action in the conformal gauge is
 \begin{align}
 \label{stringaction}
 S_{\rm st}=\frac{L^2}{2\ell_s^2}\int_0^1 d\tau \int_{\sigma_0}^\infty d\sigma \left(
 r^2\partial X_\mu  \bar\partial X_\mu + \frac{\partial r \bar\partial r}{r^2}\right)
 + i\oint A
 \end{align}
 where $(\partial,\bar\partial) =(\partial_\sigma\pm i\partial_\tau) $.  The last term is the
 coupling of the boundary of the string worldsheet to the gauge field.
 The equations of motion,
 Virasoro constraints and boundary conditions are
 \begin{align}
\partial\bar\partial  r=r^3\partial X\bar\partial X+\frac{1}{r}\partial r
 \bar\partial r   ~~&,~~  \partial_a(r^2 \partial^a X_\mu)=0
  \\
 r^2(\partial X)^2+\frac{(\partial r)^2}{r^2}=0~~&,~~r^2(\bar \partial X)^2+  \frac{(\bar\partial r)^2}{r^2}=0
 \\
  X(\tau, \sigma_0)=R~\hat r(\tau) ~~&,~~ r(\tau, \sigma_0)=r_0
  \end{align}
  They have the solution
  \begin{align}\label{stringsolutions}
  X= \frac{\cosh(2\pi n\sigma_0)}{ \cosh (2\pi n \sigma )}~R~\hat r
 ~,~
  r=  r_0\frac{\tan(2\pi n\sigma_0)}{\tanh(2\pi n\sigma )}
  \end{align}
  when $\sinh(2\pi n\sigma_0)=1/Rr_0$.
  We then replace (\ref{classical}) with the on-shell string action, found by substituting
  (\ref{stringsolutions}) into (\ref{stringaction}) and using (\ref{wbosonmass}) to get
  \footnote{A formula similar to this, but with a different conclusion for the value of the critical field
  was derived in Ref.~\cite{Gorsky:2001up}.}
  \begin{align}\label{classical2}
  S_{\rm cl(2)}=n\left[\sqrt{(2\pi  m R)^2+\lambda}-\sqrt{\lambda}\right]-\tfrac{1}{2}(2\pi n)ER^2
  \end{align}
  This expression should be accurate when $mR$ is large and when $\lambda$ is large.  It
  reproduces (\ref{classical}),
  corrected by the Wilson loop term $-n\sqrt{\lambda}$ in the limit where $mR >> \sqrt{\lambda}$.
   The radius should now be fixed to an extremum  of (\ref{classical2}),
   \begin{align}R=
  \frac{1}{2\pi m}\sqrt{ \left(\frac{2\pi m^2}{E}\right)^2-\lambda}
  \label{R}\end{align}
 There is a critical value of
  the electric field where this radius shrinks to zero, given by the value of the
  critical field $E_c$ in (\ref{emax0}). The classical
  action
  \begin{equation}
  S_{\rm cl(2)}=\tfrac{n\sqrt{\lambda}}{2}\left(\sqrt{\tfrac{E_c}{E}}-\sqrt{\tfrac{E}{E_c}}\right)^2
  \end{equation}
  also vanishes when $E$ approaches $E_c$ and the summation over $n$ is unsuppressed. Moreover, it agrees with
  the Schwinger result in the weak field $E<<E_c$ limit.

\begin{figure}[t]
\centerline{\includegraphics[scale=0.35]{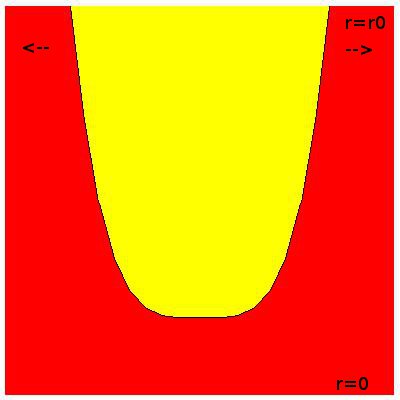}}
\caption{\label{astr}\small String profile.  The endpoints of the string
at $r=r_0$ have constant proper acceleration in opposite directions.  \label{ads2}}
\end{figure}

The worldsheet in (\ref{stringsolutions}) can be continued to Lorentzian signature  where it
is the locus of $-t^2+x^2+\tfrac{1}{r}^2=R^2+\tfrac{1}{r_0^2}$ with $r\leq r_0$, that is
part of $AdS_2$. At any fixed time, $t $, the profile of the string is
\begin{equation}
x(t,r)=\pm \sqrt{t^2+R^2+\tfrac{1}{r_0^2}-\tfrac{1}{r^2} }
\label{stringdrag}\end{equation}
and is depicted in figure \ref{astr}.
The endpoints on the probe brane at $r=r_0$ are the
position of the particle and the anti-particle $\pm\sqrt{t^2+R^2}$.
At $t=0$, they are separated by
 a distance $2R $.  After the initial time, the particle  and antiparticle
 follow  trajectories with constant proper
acceleration of magnitude $a=\frac{1}{R}$ and in opposite directions.  If they were simple charged particles
with mass $m$,
the electric field would give them proper acceleration $a=\frac{E}{m}$, agreeing with the
result $R=\frac{m}{E}$ for the radius that was found by extremizing the action (\ref{classical}).
When we extremize the stringy action (\ref{classical2}) to get (\ref{R}), the radius has decreased,
so the proper acceleration is greater, $a=\frac{E/m}{\sqrt{1-(E/E_c)^2}}$. The endpoint of the
string in a strong electric field seems to have less inertia than a particle would have.  Remember that a particle is a string
which hangs from the probe brane
to the Poincar\'{e} horizon, whereas the appropriate string for our problem here, shown
in figure \ref{astr},
never reaches the Poincar\`{e} horizon, rather it lags behind the endpoint and it joins
 with the string of the anti-particle.   This join persists, regardless of the
 separation. (There is no Gross-Ooguri phase transition \cite{grossooguri} in this case.)  This joining of the
 string is responsible for the classical action in (\ref{classical2})
which is slightly smaller an the analogous one for a relativistic particle
given in (\ref{classical}).  In the field theory language, the effect can be seen as coming from the $\lambda$-dependence of
the action and it reflects the negative
Coulomb interaction energy of the particle and anti-particle, which persists at strong coupling, and was the basis for our argument leading to eq.~(\ref{emax00}).

\subsection*{Acknowledgments}
We  thank J.~Ambj\o{}rn, J.~Davis, M.~Heller, V.~Hubeny, R.~Myers, M.~Rangamani,
M.~Rozali, A.~Sever, M. van Raamsdonk and P.~Vieira
for discussions. K.Z. acknowledges the Perimeter
Institute for hospitality and financial support of the Swedish Research Council
under contract 621-2007-4177,  ANF-a grant 09-02-91005,
  RFFI grant 10-02-01315, and the Ministry of Education and Science of the Russian Federation
under contract 14.740.11.0347. G.W.S.~ is supported by NSERC of Canada and acknowledges
the Aspen Center for Physics,  KITP Santa Barbara, Galileo Galilei Institute
and Nordita, where parts of this work were completed. Work done at KITP is supported in part by the National
Science Foundation under Grant No.~NSFPHY05-51164 and in part by DARPA under Grant No.
HR0011-09-1-0015 and by the National Science Foundation under Grant
No. PHY05-51164.


\begin{thebibliography}{99}

\bibitem{Fradkin:1985ys}
  E.~S.~Fradkin, A.~Tseytlin,
  Nucl.~Phys.~B{\bf 261}, 1 (1985).

\bibitem{Bachas:1992bh}
  C.~Bachas and M.~Porrati,
  Phys.\ Lett.\  B {\bf 296}, 77 (1992)
  [arXiv:hep-th/9209032].

\bibitem{Schwinger:1951nm}
  J.~S.~Schwinger,
  Phys.\ Rev.\  {\bf 82}, 664 (1951).

\bibitem{Affleck:1981bma}
  I.~K.~Affleck, O.~Alvarez and N.~S.~Manton,
  Nucl.\ Phys.\  B {\bf 197}, 509 (1982).


\bibitem{Rey:1998ik}
  S.~J.~Rey and J.~T.~Yee,
  Eur.\ Phys.\ J.\  C {\bf 22}, 379 (2001)
  [arXiv:hep-th/9803001].

\bibitem{Maldacena:1998im}
  J.~M.~Maldacena,
  Phys.\ Rev.\ Lett.\  {\bf 80}, 4859 (1998)
  [arXiv:hep-th/9803002].


\bibitem{Karch:2002sh}
  A.~Karch and E.~Katz,
  JHEP {\bf 0206}, 043 (2002)
  [arXiv:hep-th/0205236];
\bibitem{Karch:2007pd}
  A.~Karch, A.~O'Bannon,
  JHEP {\bf 0709}, 024 (2007)
  [arXiv:0705.3870];
  M.~Ammon, T.~H.~Ngo, A.~O'Bannon,
  JHEP {\bf 0910}, 027 (2009)
  [arXiv:0908.2625].



\bibitem{Drukker:1999zq}
  N.~Drukker, D.~J.~Gross and H.~Ooguri,
  Phys.\ Rev.\  D {\bf 60}, 125006 (1999)
  [arXiv:hep-th/9904191].

\bibitem{Erickson:2000af}
  J.~K.~Erickson, G.~W.~Semenoff and K.~Zarembo,
  Nucl.\ Phys.\  B {\bf 582}, 155 (2000)
  [arXiv:hep-th/0003055].

\bibitem{Drukker:2000rr}
  N.~Drukker and D.~J.~Gross,
  J.\ Math.\ Phys.\  {\bf 42}, 2896 (2001)
  [arXiv:hep-th/0010274].

\bibitem{Pestun:2007rz}
  V.~Pestun,
  arXiv:0712.2824 [hep-th].



\bibitem{Polyakov:2000ti}
  A.~M.~Polyakov and V.~S.~Rychkov,
  Nucl.\ Phys.\  B {\bf 581}, 116 (2000)
  [arXiv:hep-th/0002106].

\bibitem{Semenoff:2004qr}
  G.~W.~Semenoff and D.~Young,
  Int.\ J.\ Mod.\ Phys.\  A {\bf 20}, 2833 (2005)
  [arXiv:hep-th/0405288].

\bibitem{Gorsky:2001up}
  A.~S.~Gorsky, K.~A.~Saraikin, K.~G.~Selivanov,
  Nucl.\ Phys.\  {\bf B628}, 270-294 (2002).
  [hep-th/0110178].


\bibitem{grossooguri}
D.~J.~Gross, H.~Ooguri, Phys.~Rev.~{\bf D58} 106002 (1998);
arXiv:hep-th/9805129v2


\end{thebibliography}

\end{document}